\documentclass[iop]{emulateapj}
\slugcomment{Accepted for publication in ApJL}

\begin{document}

\title{The Arecibo Detection of the Coolest Radio-flaring Brown Dwarf}

\author{M. Route\altaffilmark{1,2} and A. Wolszczan\altaffilmark{1,2}}

\altaffiltext{1}{Department of Astronomy and Astrophysics, The Pennsylvania State University, 525 Davey Laboratory, University Park, PA 16802, USA; mroute@astro.psu.edu, alex@astro.psu.edu}

\altaffiltext{2}{Center for Exoplanets and Habitable Worlds, The Pennsylvania State University, 525 Davey Laboratory, University Park, PA 16802, USA}

\slugcomment{Accepted for publication in ApJL}

\begin{abstract}
Radio detection provides unique means to measure and study magnetic fields of the coolest brown dwarfs. Previous radio surveys have observed quiescent and flaring emission from brown dwarfs down to spectral type L3.5, but only upper limits have been established for even cooler objects. We report the detection of sporadic, circularly polarized flares from the T6.5 dwarf, 2MASS J1047+21, with the Arecibo radio telescope at 4.75 GHz. This is by far the coolest brown dwarf yet detected at radio frequencies. The fact that such an object is capable of generating observable, coherent radio emission, despite its very low, $\sim$900 K temperature, demonstrates the feasibility of studies of brown dwarfs of the meagerly explored L, T, and Y spectral types, using radio detection as a tool.
\end{abstract}
 
 
\section{Introduction}

A phenomenological description of the magnetic activity of the lowest mass stars and brown dwarfs (hereafter ultracool dwarfs) has been constructed on the basis of several types of observations, including X-ray, H$\alpha$, and radio emission \citep{ber08a,ber08b,ber09,ber10}, as well as Zeeman broadening of spectral lines \citep{rei07,rei10}, and Zeeman-Doppler imaging of these objects \citep{don06a,don06b,mor08a,mor08b,mor10}. The data clearly show a changeover from toroidal to low-multipole, poloidal magnetic field topology in stars of spectral types later than M3, beyond which they become fully convective \citep{mor10}, and a decline of the levels of H$\alpha$ and X-ray emission for stars later than M7. In sharp contrast, the radio flux remains largely unchanged down to at least the L3.5-type \citep{ber06,ber10}. It is particularly illuminating to correlate the magnetic activity indicators, namely the X-ray, H$\alpha$, and radio emission with parameters that are likely related to the magnetic field generation by a turbulent dynamo mechanism, such as the rotational velocity, $v$, and the Rossby number, $R_{0} = P_{rot}/\tau_{c}$ ($P_{rot}$ is the stellar rotation period and $\tau_{c}$ is the convective turnaround time) \citep{noy84}. \citet{mcl12} recently carried out a study of over 200 ultracool dwarfs and showed that the X-ray and H$\alpha$ luminosities scaled to the bolometric luminosity, saturate or even supersaturate when $vsin(i) >$ 5 km s$^{-1}$ and $R_{0} < 0.01$, whereas the radio luminosity, scaled in the same manner, does not decrease beyond these limits.

These results confirm and extend the earlier findings \citep{ber01,hal07,hal08,pha07,ber09,mcl11} and demonstrate that the ultracool dwarfs, at least as cool as the L3.5 type, do maintain kilo-Gauss magnetic fields that produce energy needed to generate the observed radio emission. At the same time, decoupling of the increasingly neutral atmospheres of these objects from the magnetic fields may decrease the efficiency of the coronal heating by magnetic energy release and suppress the X-ray emission \citep{moh02,ber08b}. The stripping of stellar coronae by rapid rotation of ultracool dwarfs may provide another way to achieve the same effect \citep{ber08a,ber08b}. Similar explanations hold for the observed decline in the chromospheric H$\alpha$ emission from stars later than M7. In practical terms, these trends also imply that, because X-ray and H$\alpha$ detection of the coolest brown dwarfs is difficult, their usefulness as magnetic activity indicators becomes very limited \citep{ber10}. In addition, the rapid rotation of these objects impairs the sensitivity of Zeeman polarimetry techniques due to spectral line broadening. Consequently, it is radio detection that becomes the tool of choice to study the generation, topology, and the mechanism of energy release into neutral atmospheres of magnetic fields in the coolest dwarfs, which populate the late L and T spectral range.

We have recently initiated a 4.75 GHz monitoring program of a sample of the LT dwarfs in the declination range of the 305-m Arecibo  radio telescope.  Here, we report the discovery of sporadic bursts of polarized radio emission from the T6.5 brown dwarf 2MASS J10475385+2124234 (hereafter J1047+21). We describe the Arecibo observations of this object in Section 2. The basic properties of J1047+21 deduced from the measurements, and possible consequences of our detection are discussed in Sections 3 and 4, respectively.

\section{Observations}

J1047+21 was first detected during a search of the Two Micron All-Sky Survey (2MASS) archive for brown dwarfs \citep{bur99}.  Subsequent 2MASS photometry and near-infrared spectroscopy of J1047+21 have led to its categorization as a T6.5V dwarf \citep{bur02}. \citet{vrb04} estimated the effective temperature of J1047+21 to be about 900 K with a distance of 10.3 pc from parallax measurement. Weak H$\alpha$ emission from the brown dwarf has been measured by \citet{bur03b}. Subsequent observations of J1047+21 with the Very Large Array (VLA)  have set a 45 $\mu$Jy upper limit to its quiescent radio flux at 8.46 GHz \citep{ber06}.

J1047+21 was observed as part of our ongoing survey of a sample of the coolest brown dwarfs visible from Arecibo, in search of rapidly varying, polarized radio emission from these objects. Observations were made with the 305-m Arecibo radio telescope equipped with the 5 GHz, dual-linear polarization receiver. For pointings at low zenith angles, the antenna gain and the system temperature were $\sim$8 K Jy$^{-1}$ and $\sim$30 K, respectively. We recorded the time resolved dynamic spectra at the source position for about 2 hr of the available tracking time of the telescope. The spectra for the four Stokes parameters were sampled at 0.1 s intervals by seven 172 MHz-bandwidth, 8192-channel, FPGA-based MOCK spectrometers\footnotemark, individually tuned to give an almost continuous coverage of a 1 GHz bandpass centered at 4.75 GHz. The 1$\sigma$ sensitivity of our observations to broadband bursts from J1047+21, binned to 0.9 s resolution, was $\sim$0.15 mJy. We emphasize the benefits of using broadband, high-resolution dynamic spectra of rapidly varying stellar emissions as a means of unraveling the basic physics of these phenomena. This is the observing method of choice in solar physics \citep{bas98}, and it has been very successfully used in stellar flare detections (e.g., Osten \& Bastian 2008; Trigilio et al. 2011). 

\footnotetext[3]{ ``The Mock Spectrometer,'' available at http://www.naic.edu/~astro/mock.shtml.}

Bursts from J1047+21 were detected three times in 15 monitoring observations spread over a 13-month period, each observing session lasting 1.5-2 hr. The first burst was detected on 2010 January 6, followed by another detection on 2010 December 5, during a weeklong campaign to confirm the initial observation. Finally, yet another successful observation of a bursting radio emission from this object was made on 2011 February 7. To our knowledge, the J1047+21 observations reported here are the first presentation of radio bursts from a brown dwarf recorded as high resolution dynamic spectra over a continuous, 1 GHz band.

\begin{figure}
\centering
\includegraphics[width=0.8\textwidth,angle=0]{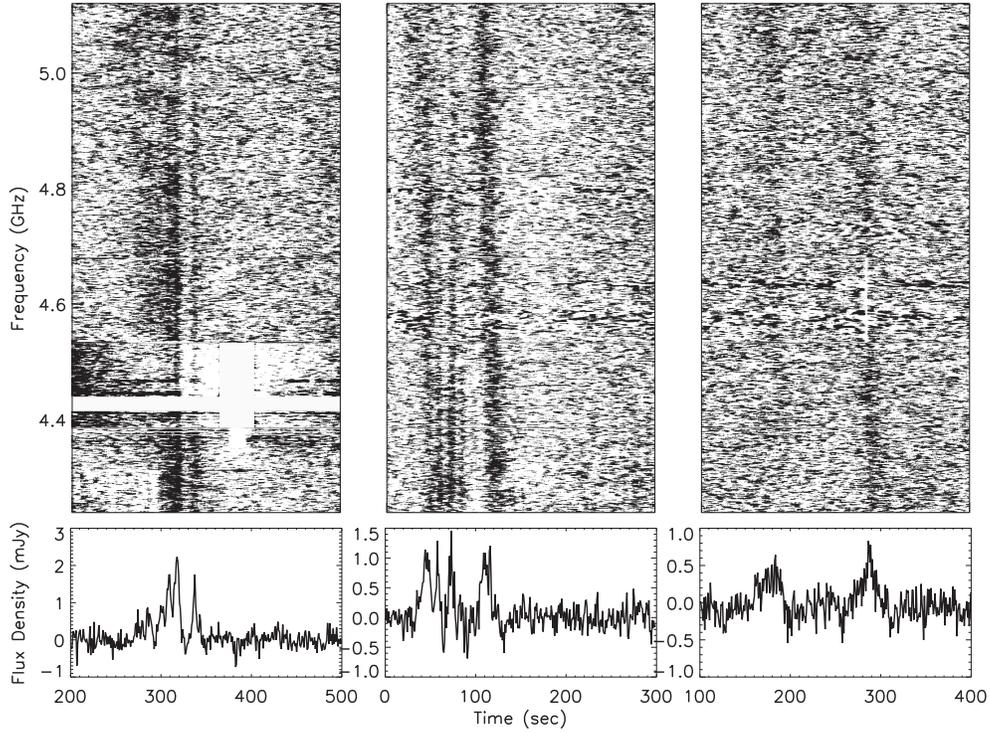}
\caption{Dynamic spectra (top) and average time profiles (bottom) of the three Stokes V bursts from J1047+21. The bursts are arranged chronologically from left to right. Darker shades of gray represent larger flux density of the right-handed circularly polarized signal. In the first burst, the data between 4.35 and 4.55 GHz are affected by exceptionally strong interference. A white vertical line around 4.6 GHz in burst three is also an artifact of interference removal. For display purposes, the dynamic spectra have been smoothed and binned to the resolution of 6 s and 2 MHz in time and frequency, respectively. The time profiles have been integrated over the 1 GHz spectrometer bandpass and binned to 0.9 s resolution.  \label{fig1}}
\end{figure}

The dynamic spectra and bandpass-averaged Stokes V burst profiles are shown in Fig. \ref{fig1}. The first two bursts have peak flux densities of 2.7$\pm$0.2 mJy and 1.3$\pm$0.2 mJy, respectively, and are highly circularly polarized (89$\pm$5$\%$ and 72$\pm$7$\%$). They have similar total widths of $\sim$100 s, and exhibit rapid, $\leq$10 s intensity variations. The third, weaker, less-polarized burst (0.8$\pm$0.2 mJy, 18$\pm$2$\%$) consists of three approximately equally spaced, $\sim$60 s peaks, two of which are shown in Fig. \ref{fig1} (the trailing, weakest peak is omitted to preserve the same display timescale for the three bursts). In this case, the observed emission could be depolarized due to propagation effects \citep{hal06,hal08}. All three bursts share a $\sim$150 to $\sim$300 MHz s$^{-1}$ drift from higher to lower frequencies with an instantaneous bandwidth of 0.5-1 GHz. The coolest extrasolar body of a substellar mass for which radio emission has been previously detected was the L3.5 dwarf 2MASS J00361617+1821104 \citep{ber02,ber05,hal08}, with an inferred temperature of 1900-2000 K \citep{vrb04}. Our discovery pushes this limit $\sim$1000 K down the temperature scale and demonstrates that brown dwarfs that are not much hotter than giant planets are still capable of generating observable radio emission.

Because Arecibo is a single, fixed-dish telescope, it has a restricted practical sensitivity to weak, quiescent emission from radio sources. This is due to a large beam width and the related confusion limitations, the presence of man-made interference, and difficulties in obtaining a precise calibration of ON-OFF-source measurements caused by the telescope aperture blockage and signal reflections from the surrounding terrain. On the other hand, the telescope is an excellent instrument for detection of rapidly varying, broadband, circularly polarized radio emission, in which case the calibration by position switching is not necessary and the signal is almost interference-free, except for occasional strong bursts that are easy to identify (Fig. \ref{fig1}). In addition, we have not found any sources of a detectable radio emission inside the $\sim$1 arcmin telescope beam centered on J1047+21. The nearest sources are two radio galaxies, FIRST J104806.0+212559 and FIRST J104807.4+212601\footnotemark, located at the respective angular distances of 3.4 and 3.7 arcmin from our target, which places them well beyond the first, 2 arcmin sidelobe of the telescope beam. For these reasons, and because of the broadband nature, circular polarization, and the shared frequency drift of the three bursts detected from the direction of J1047+21, we find it very unlikely that they could originate from a source other than the brown dwarf itself.

\footnotetext[4]{ ``The VLA FIRST Survey,'' available at http://sundog.stsci.edu/index.html.}

Given the existing evidence that, in some cases, the observed bursts of radio emission from brown dwarfs are periodic in synchronism with their rotation (e.g., Hallinan et al. 2007, 2008; Berger et al. 2009; McLean et al. 2011), it is possible that J1047+21 belongs to the same category. Obviously, additional detections of bursts from this source are needed to unambiguously determine their time dependence.

\section{Characteristics of J1047+21}

Properties of the 4.75 GHz radio emission from J1047+21 are broadly similar to those observed in bursts from other ultracool dwarfs \citep{ber01,hal07,hal08,ber09,mcl11}, the Sun \citep{che06,whi07}, and the solar system planets \citep{ark09,lam10,mut10}. A closer examination of sufficiently high signal-to-noise sections of the first two bursts reveals spikes that last no longer than $\sim$0.5 s. The presence of rapid intensity fluctuations in the data suggests that the shortest observable timescales of these variations may be limited by our 0.1-s sampling. In this case, the light travel time argument restricts the source size, $R$, to $\leq$0.4  of the Jupiter radius, $R_{J}$. Given the measured flux densities of the bursts, $F_{\nu}$, the emission frequency, $\nu$, and the distance to source, $d$, this yields estimates of their brightness temperatures, $T_{b} = 2 \times 10^{9} F_{\nu,mJy}\nu^{-2}_{GHz} d^{2}_{pc}(R/R_{J})^{-2}$ K, to be $>$10$^{11}$ K for the first two bursts and $>$5$\times$10$^{10}$  K for the weaker third one. As in the case of the highly polarized bursts from other ultracool dwarfs, these temperatures indicate that the observed emissions come from a coherent source powered by the electron cyclotron maser instability (ECMI) \citep{tre06}. Under this assumption, the magnetic field strength, $B$, of J1047+21 can be estimated from $\nu_{c} = 2.8 \times 10^{6} B$ (Gauss), where $\nu_{c}$, the local electron cyclotron frequency, is taken to be equal to 4.75 $\times$ 10$^{9}$ Hz, the center frequency of our observations. The resulting value of $B = 1.7$ kG is similar to those obtained for other ultracool dwarfs, for which coherent radio emission has been observed \citep{ber01,ber09,hal07,hal08,mcl11}.

Because the ECMI must be generated well above the local plasma frequency, $\nu_{p} \approx 9 \times 10^{3} n_{e}^{1/2}$, one can place an upper limit on the electron density in the emission region, $n_{e}$ (cm$^{-3}$), by requiring that $\nu_{c} > \nu_{p}$. This condition should be easy to satisfy in the very cool, largely neutral atmosphere of a T6.5 dwarf. The resulting limit, $n_{e} < 3 \times$ 10$^{11}$ cm$^{-3}$, is again similar to electron densities computed from radio detections of other dwarfs \citep{ber01,ber08b,ber06}. In addition, it provides a useful means to examine the nature of a negative frequency drift observed in the three bursts from J1047+21. In principle, such a drift could be caused by dispersion of a radio burst by free electrons along the line of sight, which would temporally broaden it by $\Delta t$ (ms) over the bandwidth $\nu_{2}-\nu_{1}$ (MHz) according to the relationship: 
\begin{equation} \Delta t = 4.15 \times 10^{6} (\frac{1} {\nu_{1}^{2}} - \frac{1} {\nu_{2}^{2}})\int_{0}^{d} n_{e} dl,\end{equation}
where the line-of-sight integral of $n_{e}$ is the so-called dispersion measure expressed in units of cm$^{-3}$  pc \citep{lor05}. Assuming that $d = R_{J}$, yields $\Delta t < 54$ ms, which is less than the 0.1- s sampling interval of the spectra. Furthermore, as the average electron density in the interstellar medium amounts to  $\sim$0.03 cm$^{-3}$ \citep{lyn06}, and the distance to J1047+21 is only 10.3 pc, the interstellar electrons make a negligible, $<$24 $\mu$s contribution to the observed temporal smearing of bursts by the frequency drift. Similarly, as discussed by \citet{ost08}, diffraction and refraction effects on the signal, induced by electron density irregularities along the line of sight, can be safely ignored. Consequently, a remaining, logical alternative is that the persistent negative frequency drift of bursts observed in J1047+21 is due to an upward propagation of the emitter in the weakening magnetic field.

The observed frequency drift of the three bursts is qualitatively similar to the drifts observed in the terrestrial auroral kilometric radiation (AKR, e.g., Zarka 1998), solar bursts associated with the coronal mass ejections (CME, e.g., Hudson et al. 2001), and bursts from some stellar sources (e.g., Osten \& Bastian 2008). Assuming the magnetic field geometry of the star and the ECMI origin of its radio bursts, one can constrain the size of the emitting region independently of the light travel time argument used above. If the magnetic field can be represented by a simple dipole, $B=B_{0}(a/r)^{3}$, where $a$ is the scale size of the field, and the observed frequency drift, $\dot{\nu}$, represents the changing local value of $\nu_{c}$ as the emitting region moves along the magnetic field lines, its velocity $v$ can be approximately expressed as $v \approx 47 aB_{0}^{1/3}\dot{\nu}\nu_{c}^{-4/3}$ \citep{ost08}. Substituting the observed values of $\nu$ = 4.75 GHz for $\nu_{c}$,  $\dot{\nu}$= 200 MHz s$^{-1}$, and $B_{0}$ = 1.7 kG, and assuming the propagation speed of the emitter to be similar to that observed in the AKR or CME bursts (300-1000 km s$^{-1}$; e.g., Treumann 2006; Hudson et al. 2001), one obtains $a \approx 0.3R_{J}-1R_{J}$, in rough agreement with the light travel time estimate.

\section{Discussion}

\begin{figure}
\centering
\includegraphics[width=0.8\textwidth,angle=0]{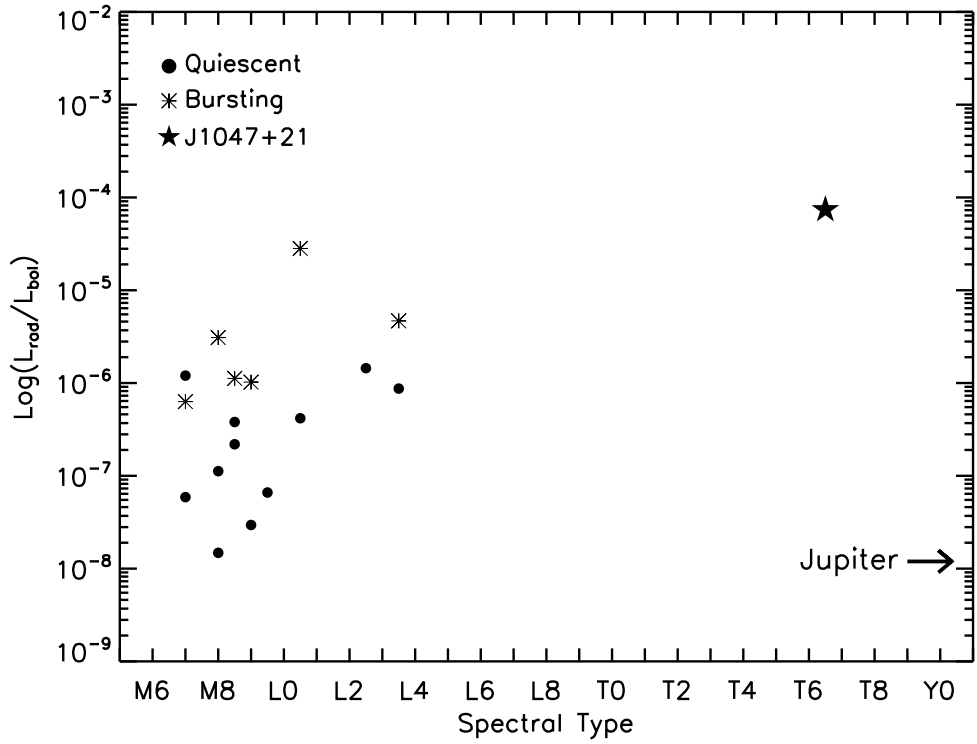}
\caption{Radio detections of ultracool dwarfs of spectral types later than M7 displayed as ratios of radio luminosity to bolometric luminosity (McLean et al. 2012 and references therein). The luminosity of J1047+21 is the average of the two high circular polarization bursts.  For comparison, Jupiter's ratio of radio to bolometric luminosity \citep{gui05,laz07} is marked with an arrow.  \label{fig2}}
\end{figure}

The status of radio observations of ultracool dwarfs, cooler than M6, is illustrated in Fig. \ref{fig2}. Our detection of flares from J1047+21 indicates that brown dwarfs remain intermittently active as radio emitters at least down to the spectral type T7. This conclusion is also consistent with a marginal detection of the H$\alpha$ emission from J1047+21 \citep{bur03b}, which confirms that conditions similar to those existing in the chromospheres of the hotter, radio detectable stars must arise in this object. Because the atmosphere of a $\sim$900 K dwarf is largely neutral, it is plausible that the necessary energy is generated by interior magnetic stresses and then transported to the upper atmosphere to give rise to the observed flaring behavior as it dissipates \citep{moh02}. The persistence of circularly polarized radio flares from J1047+21 over a period exceeding 1 yr indicates that the object's magnetic field remains stable and the flare generation mechanism continues to operate on timescales that are at least that long. The similarity of the observed and derived characteristics of J1047+21 to those of hotter brown dwarfs implies that radio detections of these objects at gigahertz frequencies should be possible over the entire LT range, possibly extending into the emerging Y-spectral type \citep{kir99,luh12}.

In principle, the observed radio activity of J1047+21 could be related to its possible binarity, as observed in RS CVn or ML-dwarf binaries \citep{dem93,sil06,ber09}. However, the T-dwarf survey by \citet{bur03a} has not detected a binary companion to J1047+21 out to $\sim$4 AU separation, and $\ge$0.4 mass ratio. Although this result does not eliminate a more relevant possibility that J1047+21 has a very close companion, it appears that, in absence of any spectroscopic or photometric evidence of binarity of this object, it is reasonable to assume its solitary character.

Another implication of our result is the possibility that young, low mass brown dwarfs and massive exoplanets, which occupy the gray area between these two types of objects, could be detectable at gigahertz frequencies. For example, calculations by \citet{rc10} indicate that a rapidly rotating, 10$^{7}$ year old, 13 M$_{J}$ planet could generate a $\sim$1 kG magnetic field. As an illustration, among the existing hot, young planetary systems, the HR 8799 planets \citep{mar08,mar10} could in principle be detectable around 1.4 GHz at $>$0.1 mJy level assuming ECMI produced emission with source sizes and brightness temperatures similar to those discussed here. This would offer an alternative to the hitherto unsuccessful efforts to detect much older, Jupiter-level magnetic field exoplanets at low radio frequencies \citep{laz09,lec09}.

\section{Acknowledgements}

We thank Phil Perillat for his assistance with observations and data reduction. MR acknowledges support from the Center for Exoplanets and Habitable Worlds and the Zaccheus Daniel Fellowship. The Center for Exoplanets and Habitable Worlds is supported by the Pennsylvania State University, and the Eberly College of Science. The Arecibo Observatory is operated by SRI International under a cooperative agreement with the National Science Foundation (AST-1100968), and in alliance with Ana G. M\'{e}ndez-Universidad Metropolitana, and the Universities Space Research Association.  

%\clearpage

\clearpage
\newpage

\end{document}